\documentclass[prb,aps,twocolumn]{revtex4}
\usepackage{bm}
\usepackage{graphicx}
\usepackage{subfigure}
\usepackage{amsmath}
\begin{document} 
\title{Intramolecular vibrational energy redistribution in 
DCO ($\widetilde{X} ^{2}A'$): Classical-Quantum correspondence, dynamical
assignments of highly excited states, and phase space transport}
\author{Aravindan Semparithi and Srihari Keshavamurthy}
\affiliation{Department of Chemistry, Indian Institute
of Technology, Kanpur, India 208 016}
\date{\today}
\begin{abstract}
Intramolecular dynamics of highly excited
DCO ($\widetilde{X} ^{2}A'$) is studied from a classical-quantum
correspondence perspective using the effective spectroscopic Hamiltonian
proposed recently by Tr\"{o}llsch and 
Temps (Z. Phys. Chem. {\bf 215}, 207 (2001)).
This work focuses on the polyads $P=3$ and $P=4$ corresponding to excitation
energies $E_{v} \approx$ 5100 cm$^{-1}$ and 7000 cm$^{-1}$ respectively.
The majority of states belonging to these polyads are dynamically assigned,
despite extensive stochasticity in the classical phase space, using
the recently proposed technique of level velocities. 
A wavelet based time-frequency analysis is
used to reveal the nature of phase
space transport and the relevant
dynamical bottlenecks. The local frequency analysis clearly
illustrates the existence of mode-specific IVR dynamics
{\it i.e.,} differing nature of the IVR dynamics ensuing from
CO stretch and the DCO bend bright states. In addition the
role of the weak Fermi resonance involving the CO stretch and DCO bend
modes is investigated.
A key feature of the present work is that the techniques utilized for
the analysis {\it i.e.,} parametric variations and local frequency analysis
are not limited by the dimensionality of the system.
This study, thus, explores the potential for understanding IVR 
in larger molecules from both
time domain and frequency domain perspectives.

\end{abstract}
\maketitle

\nopagebreak

\section{Introduction}

Substantial progress has been made over the past few decades in
understanding the nature of energy flow in highly excited
molecules\cite{bj68,sar81,tu91,leh94,nes96,mg98,mg20,kesk20}.  
At higher levels of excitation the usual
rigid rotor-normal mode description breaks down and the modes 
get strongly coupled. The resulting spectral complexity in terms of the
splittings and intensities encode the flow of energy through the molecule.
This phenomenon of intramolecular vibrational energy
redistribution (IVR) can lead to subtle but
important deviations from the predictions 
of statistical theories like RRKM\cite{bha} and thus
profound consequences for the rates
of chemical reactions\cite{mgtca}.
The mode mixings in general have classical as well as quantum origins. Much of
the classical effects arise from vibrational anharmonicities, coriolis
and centrifugal distortion effects. Possible
quantum contributions\cite{hu84,stu93,la79,dav81,ejh95,ks03} manifest
as indirect state-to-state explorations
involving a sequence of intermediate, off-resonance 
virtual states (``vibrational superexchange'').
In molecules with symmetry the phenomenon of dynamical tunneling
provides a quantum route to mode mixing\cite{dav81,ejh95,ks03}. 
Theoretically it is crucial to
establish and understand IVR pathways from time dependent as well
as time independent perspectives. 
RRKM theory has been quite successful\cite{bha} 
but evidence is mounting
for a number of reactions
where the intramolecular dynamics is 
inherently nonstatistical\cite{bha,mgtca}.
Accounting for such nonstatistical dynamics is theoretically
challenging and crucial for control of reactions.

In this work our aim is to understand IVR in the deuterated formyl radical
DCO. 
Our choice of the system is primarily due to the
considerable experimental\cite{dcoexpt} and
theoretical\cite{keller,stam,renth,jung} work on IVR 
and mode specificity of the unimolecular decay
of highly vibrationally excited DCO. 
In contrast to HCO the deuterated
analog involves a strong accidental $1:1:2$ resonance between the
vibrational frequencies.
The experiments and
analysis of Renth {\it et al} using a effective Hamiltonian clearly shows that
the unimolecular decay in DCO is IVR limited\cite{renth}. 
Although DCO has three vibrational modes, $m$ (DC stretch), $n$ (CO stretch),
$b$ (bend), the coupling structure 
of the effective Hamiltonian\cite{effham} yields
a conserved quantity called as the polyad number $P=m+n+b/2$. Experimentally
the existence of $P$ is revealed by bunching of vibrational levels as
evident in the SEP spectrum of DCO\cite{dcoexpt}.

As a small molecule exhibiting nonstatistical intramolecular dynamics
DCO provides an ideal system to test the various theoretical approaches
for studying highly excited molecules. Impressive advances have been made
towards fully quantum investigation of the IVR dynamics in
large systems\cite{mg20,mgtca,wyatt,stuche,child}. 
There is little doubt that DCO can be subjected to
such exact treatments which are important in their own right. 
However there are strong reasons\cite{cqc} to believe that
powerful insights into IVR can be gained by approaches
based on classical-quantum
correspondence. 
For example, Jung {\it et al} have provided dynamical assignments
of the highly excited states of DCO based on certain key periodic
orbits in phase space\cite{jung}.
A key to the success of the many classical-quantum
correspondence 
studies\cite{cqc,sib,kel,ks,kscpl,jac,ishi,jung2,mart,ccm,kma} 
is that the understanding of IVR is firmly
rooted in the invariant structures inhabiting the underlying phase space.
This provides a very clear dynamical picture of IVR in molecules
and can serve as a baseline for estimating the important quantum
routes to mode mixings.

Currently one of the main obstacles in extending such useful classical-quantum
correspondence studies is the issue of dimesionality. 
Despite the
existence of potentially $3N-6$ coupled vibrational modes it is now well
understood that not all of the modes are coupled strongly\cite{mg20,mgtca}.
Although such hierarchicality of the couplings 
leads to much smaller dimensionalities of the `IVR space' it is
possible that 
for large enough molecules or smaller molecules at higher levels
of excitation the effective dimesionality of the system can be
larger than three. In such instances the dimensionality of the system
poses both technical as well as conceptual problems in the context
of classical-quantum correspondence studies. The technical problems are
well known and have
to do with the inability to visualize Poincar\'{e} surface of sections
or the quantum states and, computation of the periodic orbits or higher
dimensional tori. Thus the knowledge of the
global phase space structure along with its
influence on the quantum states
is at present difficult to obtain for systems with three or more degrees of
freedom. 
For more detailed discussions on the dimensionality issues, technical
and conceptual, we refer the
reader to the earlier works\cite{ardif,cinst,ks,gill,beg}.

One of the central goals of our research is to
come up with techniques that would allow a detailed
classical-quantum correspondence study of IVR in molecules
with three or greater degrees of freedom and in this context
two approaches will be of particular interest. 
The first one is the promising technique of local frequency analysis (LFA)
put forward nearly two decades ago by Martens, Davis 
and Ezra\cite{lfa1}. 
In their pioneering work\cite{lfa1,lfa2,bru,dav} 
on energy flow in the OCS molecule 
it was shown that LFA was capable of
highlighting the role of various phase space structures including
partial barriers and resonance zones. 
The important point is that although motion may be irregular
when viewed over long time scales (several picoseconds), quantities
such as frequencies can be constant over times corresponding to
many vibrational periods.  This gives rise to long time correlations and
the notion of dynamically significant phase space regions\cite{lfa1,lfa2}.
Since then very little work has
been done in utilizing this technique for studying energy transport
in multimode molecules. Interestingly much more efforts have been made in
celestial mechanics\cite{lask} with fewer 
applications in atomic physics\cite{hyd} and molecular systems\cite{linc}. 
Recently Arevalo and Wiggins\cite{are1,are2} have
proposed a technique for performing
time-frequency analysis (LFA) which 
include strongly chaotic trajectories as well
by using wavelet analysis. Applications to various systems demonstrated
the advantages of a wavelet based LFA in understanding
phase space transport\cite{are2}. 

A second technique\cite{par1}
for analyzing the quantum states, based on
parametric variations, was recently proposed by us and applied to a 2-mode
model Hamiltonian. It was shown that variations of the eigenvalues
with appropriate parameters {\it i.e.,} level velocities were strongly
correlated to the important classical phase space structures organizing
the quantum states. In the case of mixed and near-integrable systems
the magnitudes and relative signs of the level velocties clearly
indicated\cite{par1} the dynamical region of phase space associated with a
given eigenstate thereby providing a dynamical assignment of the states.
The technique based on parametric variations is dimensionality independent,
like the local frequency analysis mentioned before, and more importantly
basis invariance is guaranteed. Moreover parametric variations can
be used to identify localization, resulting
in deviations from random matrix theory (RMT) predictions, 
even in strongly chaotic systems\cite{corr1}.
The full utility of parametric techniques is yet to be explored,
particularly in the context of IVR\cite{corr2}, and there do exist detailed
semiclassical theories in the (near)integrable\cite{corr3} 
and chaotic situations\cite{corr1}.

Our system of choice DCO is effectively two dimensional due to the
existence of the polyad $P$. Thus the intramolecular dynamics can be
investigated in great detail using the existing methods. However
DCO provides a stringent test for the dimensionality independent
techniques mentioned above since there is significant chaos in the
classical phase space for the lowest energies in $P=3$ and $4$.
This paper is organized as follows:
In section II the main features of
the effective Hamiltonian and its classical limit are
presented. This is followed, in section III, by a dynamical assignment of
states in $P=3,4$ using the level velocities. 
It will be shown that a dynamical assignment based on level velocities,
in conjunction with the inverse participation
ratios\cite{ipr} (IPR), can be given despite the presence
of significant chaos.
In section IV the time dependent dynamics of DCO are analyzed using
local frequency analysis. This reveals the global picture of phase space
transport and highlights chaotic trajectories undergoing rapid frequency
jumps mainly between the stretch-stretch 
$1:1$ resonance and the DC stretch-bend
$2:1$ Fermi resonance. Moreover it is
seen that chaotic trajectories can be trapped in resonances
for many vibrational periods. 
This is related\cite{stick} to 
the 'stickiness' of the chaotic trajectories and
is of importance to the IVR dynamics in the molecule.
The local frequency analysis also highlights the
nature of the CO stretch-bend $2:1$ Fermi resonance which is seen to play
a role in the IVR for short times. Section V concludes. 

\section{Effective Hamiltonian for DCO}

The spectroscopic
Hamiltonian, proposed recently by Tr\"{o}llsch and Temps\cite{effham}, 
was obtained through a fit
to the experimental DF and SEP
spectra. The relevant matrix elements are
\begin{subequations}
\begin{eqnarray}
H_{mnb;mnb} &=& \sum_{j} \omega_{j} \left(j+
\frac{1}{2}\right)  \\
&+& \sum_{jk} x_{jk} \left(j+\frac{1}{2}\right)
\left(k+\frac{1}{2}\right) \nonumber \\
H_{mnb;m-1,n+1,b} &=& \frac{\lambda_{1}}{2}
(m(n+1))^{1/2}  \\
&\times& \left[1+\lambda_{1}'m+\lambda_{1}''(n+1)+
\lambda_{1}'''\left(b+\frac{1}{2}\right)\right] \nonumber \\
H_{mnb;m-2,n+2,b} &=& \frac{\lambda_{2}}{2}
[m(m-1)(n+1)(n+2)]^{1/2} \\
H_{mnb;m-1,n,b+2} &=& \frac{k_{mbb}}{2}
\left[\frac{1}{2}m(b+1)(b+2)\right]^{1/2} \\
H_{mnb;m,n-1,b+2} &=& \frac{k_{nbb}}{2}
\left[\frac{1}{2}n(b+1)(b+2)\right]^{1/2}
\end{eqnarray}
\end{subequations}
with $j=m,n,b$ denoting the zeroth order
DC stretch, the CO stretch, and the bend
modes respectively.
The coupling terms $\lambda_{1}$ and $\lambda_{2}$ represent a $1:1$
and a Darling-Dennison $2:2$ resonance between the stretch modes respectively.
The Fermi resonant couplings $k_{m(n)bb}$ correspond to a $1:2$ resonance
between the DC (CO) stretch and the bend modes.
The effective Hamiltonian $\widehat{H}$ commutes with the polyad
$P_{Q} = m+n+b/2$ and hence block diagonal in $P_{Q}$.
A given $P_{Q}$ block has $(P_{Q}+1)(P_{Q}+2)/2$ states for integer polyads
and the states are labelled as $P_{Q j}$ in order of decreasing energy.
Note that the above Hamiltonian is able to 
approximately reproduce the
experimental DF spectra for polyad values upto
$P_{Q}=5$ since $P_{Q}=6$ was the
highest polyad considered in the fitting procedure. Consequently the current
work focuses on the polyads $P_{Q}=3,4$ as examples. The techniques used
for the analysis in this work are, however, applicable to higher polyads
as well. In what follows we will adopt a shorthand notation for
the two main resonances, $1:1$ DC stretch-CO stretch 
and $1:2$ DC stretch-DCO bend,
denoting them as $mn$ and $mbb$ respectively.

The classical limit Hamiltonian can be obtained using the Heisenberg
correspondence 
\begin{equation}
\widehat{a}_{j}^{\dagger} \leftrightarrow I_{j}^{1/2} e^{i\phi_{j}} \,\,\,\,\,
\widehat{a}_{j} \leftrightarrow I_{j}^{1/2} e^{-i\phi_{j}}
\end{equation}
between creation-annhilation operators and the action-angle
variables. The resulting classical Hamiltonian is 
\begin{eqnarray}
H &=& \sum_{j}\omega_{j}I_{j} + \sum_{jk} x_{jk} I_{j} I_{k} \\
&+& \Lambda_{1} (I_{m} I_{n})^{1/2}
\cos(\phi_{m}-\phi_{n})+\lambda_{2} I_{m} I_{n} 
\cos(2\phi_{m}-2\phi_{n}) \nonumber \\
&+& \frac{I_{b}}{\sqrt{2}}[ k_{mbb} I_{m}^{1/2} 
\cos(\phi_{m}-2\phi_{b})
+ k_{nbb} I_{n}^{1/2} 
\cos(\phi_{n}-2\phi_{b})] \nonumber
\end{eqnarray}
with $\Lambda_{1} \equiv \lambda_{1}[1+\lambda_{1}'I_{m}+\lambda_{1}''I_{n}+
\lambda_{1}'''I_{b}]$.
As expected the zeroth order Hamiltonian $H_{0}$ is a function of
the actions alone and the various couplings manifest themselves
as anharmonic resonances. The quantum zero point energy $E_{0}$ has been
subtracted from the classical Hamiltonian for consistency.
The classical system is effectively a two degree of freedom system due
to the existence of the conserved quantity $P_{c}=2(I_{m}+I_{n})+I_{b}$
which is the analog of the quantum polyad. As a result the classical dynamics
can be studied in a four dimensional reduced phase space. 
The reduction can be obtained by performing a canonical transformation via
the generating function:
\begin{equation}
F = (\phi_{m}-2\phi_{b})J_{m} + (\phi_{n}-2\phi_{b})J_{n} + \phi_{b}P_{c}
\end{equation}
The new action-angle variables are related to the old ones, 
$J_{m}=I_{m},J_{n}=I_{n},P_{c}=2(I_{m}+I_{n})+I_{b}$, and
$\psi_{m}=\phi_{m}-2\phi_{b},\psi_{n}=\phi_{n}-2\phi_{b},\theta=\phi_{b}$.
The resulting Hamiltonian\cite{jung} is independent
of the angle variable $\theta$ which is conjugate to $P_{c}$ and
hence an effectively two-dimensional system.

It is possible to perform a Chirikov\cite{chiri} 
analysis of the classical Hamiltonian
and obtain a resonance template in the state space\cite{ks}. 
However it is seen that in this case the Chirikov analysis, which is
based on the zeroth order $H_{0}$, is not very suitable. This is mainly
due to the reduced bend anharmonicity of DCO. Despite this limitation
useful qualitative information can be obtained regarding the approximate
location of the resonance zones and overlaps. For instance, such an
analysis predicts that the CO stretch-bend Fermi resonance ($k_{nbb}$) 
does not show up in the state space for most of the polyads.
Thus one anticipates the IVR dynamics in DCO to be dominated by
the stretch-stretch $1:1$ and DC stretch-bend $2:1$ Fermi resonances.
This observation is in agreement with earlier works\cite{renth,jung}. 
In what follows we will
show the resonance template as a guideline and the local frequency analysis
of the dynamics establishes the significant deviations from the zeroth
order picture due to strong couplings.

\section{Dynamical assignments of states in $P=3$ and $P=4$}

Considerable work has been done in the last few years to understand
and possibly assign highly excited states by invoking the 
classical-quantum correspondence principle\cite{cqc}. 
Every single method, without
exception, is based on identifying a dominant classical phase space
structure that organizes the eigenstates in a given energy range.
Indeed as DCO is effectively two dimensional due to the existence of
the polyad it is possible to use any one of the several techniques
proposed in the literature. 
However our objective in this work is to
use techniques that do not explicitly rely on determining the exact
periodic orbits and/or visualizations of the phase space and wavefunctions.
This is especially important for analyzing systems with more than
two degrees of freedom.

One such method was recently proposed by us\cite{par1} and it is based on the
method of parametric variations. In this approach the ``level velocities''
$\dot{x}_{\alpha}({\bf \tau})$ associated with an 
eigenstate $|\alpha({\bf \tau})\rangle$
are determined and analyzed. Thus if the effective, spectroscopic, Hamiltonian
has the general form $H = H_{0} + \sum_{j} \tau_{j} V_{j}$ with $V_{j}$
and $\tau_{j}$ denoting the various resonances and the corresponding
coupling strengths, then the level velocity of a specific eigenstate
associated with a coupling $V_{j}$ is obtained as:
\begin{eqnarray}
\dot{x}_{\alpha}(\tau_{j}) &\equiv& \frac{\partial E_{\alpha}({\bf \tau})}
{\partial \tau_{j}} = \langle \alpha|V_{j}|\alpha \rangle \\
&=&\sum_{{\bf n},{\bf n}'} c^{*}_{\alpha {\bf n}} c_{\alpha {\bf n}'}
V_{j;{\bf n}{\bf n}'} = 2 \sum_{P_{j}} \dot{x}_{\alpha}(\tau_{j};P_{j})
\end{eqnarray}
In the above equation ${\bf n}$ is the zeroth-order basis and
$c_{\alpha {\bf n}} = \langle {\bf n}|\alpha \rangle$. 
The level velocity of a state $|\alpha \rangle$ with respect to
$V_{j}$ can be written as a sum over the polyads $P_{j}$, corresponding
to the resonant perturbation $V_{j}$,
of the partial level velocities
$\dot{x}_{\alpha}(\tau_{j};P_{j})$.
In the integrable limit {\it i.e.,} with only $V_{j}$ present 
it is easy to see that $\dot{x}_{\alpha}(\tau_{j})=
\dot{x}_{\alpha}(\tau_{j};P_{j})$.
In the presence of other perturbations $V_{i \neq j}$ the level velocity
$\dot{x}_{\alpha}(\tau_{j})$ can have contributions from many
different $P_{j}$ indicating extensive mixing. 
Thus the level velocity essentially measures the response of the
quantum eigenvalue to a specific perturbation. 

It was established\cite{par1} in
the earlier work that $\dot{x}_{\alpha}(\tau_{j})$ is correlated to
the phase space nature of the eigenstate. For integrable, single
resonance cases the various properties of
the level velocities are well known\cite{ksjpc}. In
the case of mixed phase space systems it is still possible\cite{par1} to
dynamically assign states based on the level velocities since
the regular regions strongly influence the eigenstates.
In case of near integrable systems and, to some extent, for mixed systems
the correlation of the level velocities with the phase space
can be understood semiclassically since it is possible to write\cite{par1}
\begin{eqnarray}
\dot{x}_{\alpha}(\tau_{j}) &=& {\rm Tr}[\hat{V}_{j} |\alpha \rangle
\langle \alpha|] \\
&\approx& \frac{1}{(2 \pi \hbar)^{N}} \int d{\bf I} d{\bm \theta}
V_{j}^{W}({\bf I}, {\bm \theta}) W_{\alpha}({\bf I}, {\bm \theta})
\end{eqnarray}
where $V_{j}^{W}$ is the Weyl symbol of the operator $\hat{V}_{j}$ and
$W_{\alpha}$ is the Wigner function associated with the eigenstate
$|\alpha \rangle$. For near integrable systems the Wigner function
condenses on to the invariant torus\cite{mvb} thus explaining the observed
correlation. 

However for systems with considerable chaos it is not
{\it a priori} clear wether the correlation continues to hold.
The $P=3,4$ cases for DCO considered herein show extensive chaos
at low energies and hence provide a critical test for the level
velocity approach. In the extreme limit of an ergodic system
the Berry-Voros hypothesis states that\cite{mvb,vor,amod}
\begin{equation}
W_{\alpha}({\bf I},{\bm \theta}) = C \delta[E_{\alpha}-H({\bf I},{\bm \theta})]
\end{equation}
In this limit it is easy to see that $\dot{x}_{\alpha}$ is
essentially the parametric derivative of the smooth part of the
spectral staircase function\cite{amod}. 
This term is independent of the nature
of the dynamics and in particular refers to
contributions from short time orbits. Nevertheless it is well known that
even in strongly chaotic systems the periodic orbits (unstable) play
a crucial role\cite{scar}. 
Thus it is expected that in such systems the 
level velocities are influenced by the periodic orbits.
A more general expectation is that, independent of the nature of
the dynamics, closed orbits 
will influence the
level velocities. As the purpose of the current work is not to
provide a semiclassical expression for computing the level velocities,
we will briefly outline the reasons for such an expectation.
The arguments provided are not new and appear in an early work
by Eckhardt {\it et al.} where semiclassical matrix elements were
determined from periodic orbits\cite{eck}.
To begin with define a quantity
\begin{equation}
g_{V_{j}}(E) = {\rm Tr} \hat{G} \hat{V}_{j}
\end{equation}
with $\hat{G}$ being the energy Green's function and $\hat{V}_{j}$ 
is the perturbation of interest. Using the formal definition of $\hat{G}$
it is straightforward to show that
\begin{equation}
\rho_{V_{j}}(E) \equiv -\frac{1}{\pi} {\rm Im} g_{V_{j}}(E) = 
\sum_{\alpha} \dot{x}_{\alpha}(\tau_{j}) \delta(E-E_{\alpha})
\end{equation}
Thus $\rho_{V_{j}}(E)$ has poles at the eigenvalues $E_{\alpha}$ and
the residues are precisely the level velocities $\dot{x}_{\alpha}(\tau_{j})$.
The quantity $g_{V_{j}}(E)$ can be written in the
semiclassical limit as:
\begin{equation}
g_{V_{j}}(E) = \int d{\bf I} d{\bm \theta} V_{j}^{W}({\bf I},{\bm \theta})
G_{W}({\bf I},{\bm \theta};E)
\end{equation}
with $G_{W}$ being the Wigner transform of the Green's function.
The above integral can be evaluated via stationary phase approximation
and, apart from the smooth short time contribution, gives rise to
an oscillating term. The oscillating part can be evaluated for
integrable (Berry-Tabor)\cite{bt}, 
near-integrable (Almeida-Ullmo-Grinberg-Tomsovic)\cite{augt},
and chaotic (Gutzwiller)\cite{gutz} cases. In every case the oscillating part
depends on the stabilities of
the closed orbits and involves 
the classical time average of $V_{j}$ over the period of
the closed orbits\cite{childbook}. 
Since the
level velocities are extracted as residues of the oscillating part
it is reasonable to expect a strong influence of the closed orbits and
their stabilities. Note that extracting the velocities from this approach
is rather involved. 
Since it is difficult to obtain
bounds on the velocities using semiclassical techniques
in the mixed phase space case
and the effect of various bifurcations on the level
velocities is not known as of yet we will
explore the utility of the technique in dynamical assignments 
from a qualitative perspective. 
The basis for the assignments will be explicitly discussed for
the case $P=3$ since the arguments are fairly general.

\begin{figure*}[!]
\includegraphics*[width=5.0in,height=5.0in]{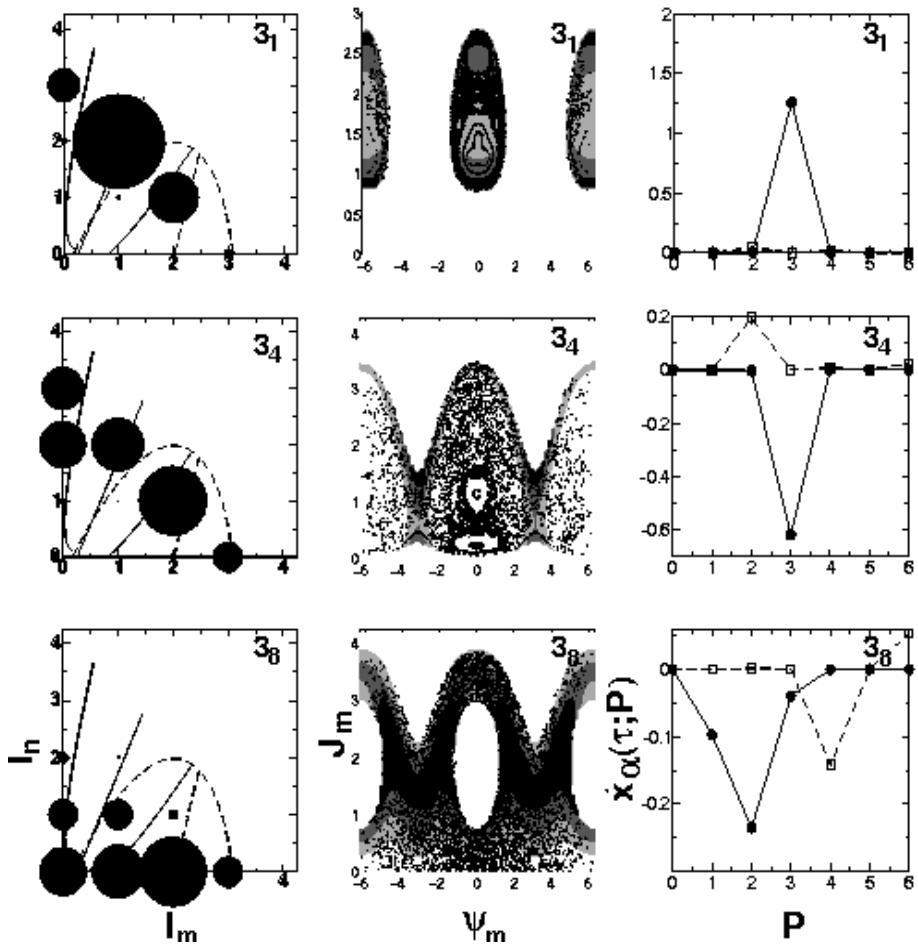}
\caption{State space representation (left panels), Husimi distributions
superposed on the surface of section (middle panel), and partial
level velocities (right panels) are shown for selected states in
$P=3$. The resonance zones computed using Chirikov approximation
are also shown in the state space. The $mn$ zone is shown as
solid lines and the $mbb$ zone is shown as the dashed lines.
Three contours of the Husimis are shown for clarity.
The states shown are 3$_{1}$ (top row), 3$_{4}$ (middle row),
and 3$_{8}$ (bottom row). Note that the Husimis of states 3$_{1}$ and
3$_{4}$ are localized around the stable and unstable regions of
the classical phase space. The Husimi of state 3$_{8}$ is delocalized
over the surface of section and the partial velocity data indicates
strong mixing. The maximum contour value for 3$_{8}$ is one third
of the maximum value for state 3$_{1}$.}
\label{fig 1}
\end{figure*}

\begin{table}
\caption{Dynamical assignments for states in $P=3$}
\label{table 1}
\begin{ruledtabular}
\begin{tabular}{cccccccc}
State & E (cm$^{-1}$) & $\dot{x}_{mn}$ & $\dot{x}_{mbb}$ & $L_{0}$ &
$L_{mn}$ & $ L_{mbb}$ & Assignment\footnote{$s, u$ and $dl$
stand for stable, unstable and distorted local respectively.
The approximate polyads $P_{mn} = m+n$ and $P_{mbb} = 2m+b$.
States involved in avoided crossings are denoted by $*$.} \\ \colrule
3$_{10}$ & 4885 & -0.09 & -1.81 & 0.55 & 0.56 & 0.91 & $P_{mbb},u$ \\
3$_{9}$  & 4935 & -0.55 & -0.40 & 0.65 & 0.80 & 0.78 & $dl$ \\
3$_{8}$  & 5032 & -0.37 & -0.09 & 0.24 & 0.33 & 0.70 & $mbb$ \\
3$_{7}$  & 5138 & -0.11 & 0.45  & 0.51 & 0.29 & 0.46 & $*, P_{mbb}$ \\
3$_{6}$  & 5177 & -0.02 & 1.09  & 0.34 & 0.33 & 0.41 &
$*, P_{mbb}$\footnote{The value of $P_{mbb}=6$ is deduced from
the partial velocity data.},$s$ \\
3$_{5}$  & 5289 & -0.31 & 0.09  & 0.33 & 0.60 & 0.39 & $P_{mn}-1$ \\
3$_{4}$  & 5354 & -0.62 & 0.23  & 0.25 & 0.69 & 0.22 & $P_{mn},u$ \\
3$_{3}$  & 5415 & 0.60  & 0.13  & 0.41 & 0.71 & 0.34 & $P_{mn}-1,s$ \\
3$_{2}$  & 5469 & 0.22  & 0.23  & 0.46 & 0.81 & 0.49 & $P_{mn}$,nom\\
3$_{1}$  & 5644 & 1.26  & 0.07  & 0.54 & 0.99 & 0.53 &
$P_{mn}$\footnote{The value of $P_{mn}=3$ is deduced from 
the partial velocity data. See Fig. \ref{fig 1}},$s$ \\
\end{tabular}
\end{ruledtabular}
\end{table}

In table \ref{table 1} we show the states in polyad $P=3$, the level
velocities with respect to the $mn$ and the $mbb$ resonances and
the IPR in three different basis. 
We start by noting that the maximum
integrable limit velocities expected (appendix 
of Ref.~\onlinecite{par1})
are $1.5$ and $2.8$ for
the $\dot{x}_{\alpha,mn}$ and $\dot{x}_{\alpha,mbb}$ respectively.
Moreover since the coupling constants $\lambda_{1}, k_{mbb} > 0$ it
is also anticipated, based on the earlier work\cite{par1}, that large positive
velocities correspond to stable (elliptic) regions of the phase space
and large negative velocities correspond to unstable (hyperbolic) regions
of the phase space. 
Consequently the highest energy state 3$_{1}$ is
clearly a $mn$ state with the corresponding Husimi localized around a 
stable region of the phase space. The high
IPR in the $mn$ basis and the small level velocity with
respect to the $mbb$ resonance clearly support such an assignment.
Thus we expect state 3$_{1}$ to be localized in the state space as well
and associate an approximate integrable limit polyad $P_{mn}$ with
the state. Our expectations are confirmed and
Fig.\ref{fig 1} shows the state space representation of 3$_{1}$
and the corresponding Husimi superposed on the surface of section.
The approximate polyad $P_{mn}$ can be extracted by looking
at the partial level velocities, 
shown in Fig.\ref{fig 1}, and one obtains $P_{mn} = 3$. 
The state 3$_{2}$ has comparable positive velocities both with respect
to $mn$ and $mbb$ {\it i.e.,} $\dot{x}_{mn}(3_{2}) \approx
\dot{x}_{mbb}(3_{2})$. This suggests that the state is influenced
by both the resonances and confirmed by inspecting the state space
representation of the state as well as the partial velocities.
However since the IPR $L_{mn}(3_{2}) > L_{mbb}(3_{2})$ and
the integrable limit $mn$ velocity is expected to be small for
this state, 
3$_{2}$ is assigned as a $mn$ state with $P_{mn} = 3$.
The next state 3$_{3}$ can be clearly assigned based on the (partial)
level velocities as well as the IPRs as a $mn$ state localized about
a stable periodic orbit with $P_{mn}=2$. State 3$_{4}$ has a negative
$mn$ velocity with large IPR in the $mn$ basis and hence it is expected
to be located in the unstable region of the phase space. The Husimi
of the state shows a typical 
seperatrix-like structure (cf. Fig.\ref{fig 1}) and hence
agrees with the velocity analysis. Based on the partial velocity data
it is straightforward to assign an approximate $P_{mn} = 3$ for 3$_{4}$.
Analogous to the state 3$_{2}$ it is possible to nominally assign
the state 3$_{5}$ as a $mn$ state with $P_{mn}=2$. 
State 3$_{10}$ has a large negative $\dot{x}_{mbb}$ and can be
clearly associated with the unstable region in phase space
with $P_{mbb}=6$. 
Again, inspection of the Husimi and the partial velocities
supports the assignment. 

\begin{figure*}[!]
\includegraphics*[width=5.0in,height=5.0in]{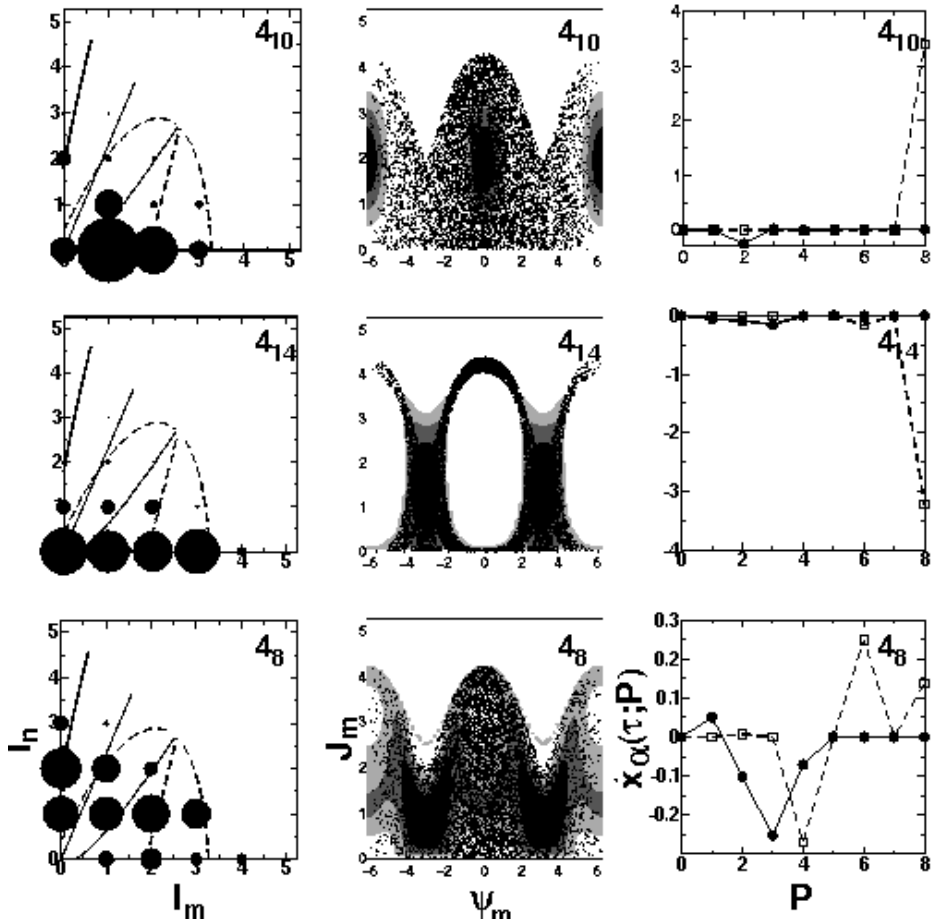}
\caption{Same as in Fig. (\ref{fig 1}) for three states in $P=4$.
The states shown are 4$_{10}$ (top row), 4$_{14}$ (middle row),
and 4$_{8}$ (bottom row). Note that the Husimis of states 4$_{10}$ and
4$_{14}$ are localized around the stable and unstable regions of
the classical phase space. The Husimi of state 4$_{8}$ is delocalized
over the surface of section and the partial velocity data indicates
strong mixing.}
\label{fig 2}
\end{figure*}

We now discuss states that are either difficult to assign or
unassignable. The particularly strong mixings in these states can
arise due to avoided crossings and/or strong overlap of the
various resonances zones.  
The notoriety of multistate avoided crossings in the process of
assignments are well known and one expects them in sufficiently
nonintegrable systems.
State 3$_{6}$
has a large positive $mbb$ velocity and is expected to be an $mbb$
state localized about a stable periodic orbit. However the velocity
is about half of the maximum expected in the integrable limit and the
IPRs are about the same in all three basis. 
The partial velocities clearly show that 3$_{6}$ can be assigned
as a $mbb$ state with $P_{mbb}=6$. 
The state 3$_{7}$ is an interesting state since the IPR in the
zeroth order basis is larger than the IPRs in either the $mn$ or
the $mbb$ basis. Indeed the state space picture and the Husimis of
the state 3$_{7}$ rule out a clear $mn$ or $mbb$ assignment. A closer
inspection reveals that the states 3$_{6}$ and 3$_{7}$ are involved in
an avoided crossing with respect to both $\lambda_{1}$ and $k_{mbb}$.
This explains the observed lower than maximum velocity of the state 3$_{6}$.
This quantum mixing, seen in earlier studies\cite{kscpl,jung}, 
is clearly 2-state
and it is possible to demix the states yielding 
states $|\alpha (\beta) \rangle = |3_{6}\rangle \pm |3_{7} \rangle$
and analyze the velocities.
On demixing it is found that $\dot{x}_{mbb}(\alpha) = 2.2$ which
is closer to the maximum value expected for an $mbb$ state in 
polyad $P=3$. The other velocities indicate dominance of the $mbb$ coupling
and $\dot{x}_{mbb}(\beta) = -0.65$ thus leading to an assignment of
the state 3$_{7}$ as $mbb$ with $P_{mbb}=4$.

State 3$_{8}$ is a typical example of a difficult
state to assign. In Fig.\ref{fig 1} the 
state space representation, Husimi, and
the partial velocity data is shown for 3$_{8}$.
The velocity data
indicates this to be a $mn$ state whereas the IPR data supports a
$mbb$ assignment. Note that this state has the lowest IPR in the
zeroth order basis amongst the set of states in $P=3$.
The particularly high value of the IPR in the $mbb$ basis suggests
that it is quite possible for this state to have a small velocity
even in the single, $mbb$ only integrable case. Indeed on computing
the integrable limit one obtains $\dot{x}_{mbb}(3_{8}) \approx -0.13$.
Thus it is possible to assign the state 3$_{8}$ as a $mbb$ state.
Note that the partial velocity data suggests $P_{mn}=4$ but the
state space representation exhibits
$P_{mbb} = 6$. The difficulty, in part, arises from the fact that
it is quite possible for a state to have a velocity close to zero
and yet be strongly influenced by that resonance\cite{ksjpc}. 
A similar situation arises for the state 3$_{9}$ in that the
level velocities are very similar. In this case, however, the
IPRs are very similar and high in all three basis. This is a typical
signature of a distorted local mode state. In this instance the
state is influenced slightly by both the resonances leading to the
similar velocities and IPRs.

\begin{table}
\caption{Dynamical assignments for states in $P=4$}
\label{table 2}
\begin{ruledtabular}
\begin{tabular}{cccccccc}
State & E (cm$^{-1}$) & $\dot{x}_{mn}$ & $\dot{x}_{mbb}$ & $L_{0}$ &
$L_{mn}$ & $ L_{mbb}$ & Assignment\footnote{The notations
are same as in table \ref{table 1}. The values
$P_{mn}=4, P_{mbb}=8$ are deduced from the 
partial velocity data. See Fig. \ref{fig 2}.} \\ \colrule
4$_{15}$ & 6167 & -0.67 & -0.31 & 0.80 & 0.95 & 0.84 & $dl$ \\
4$_{14}$ & 6392 & -0.30 & -3.38 & 0.22 & 0.25 & 0.76 & $P_{mbb},u$ \\
4$_{13}$ & 6448 & -0.30 & -0.98 & 0.36 & 0.40 & 0.76 & $P_{mbb}$ \\
4$_{12}$ & 6593 & -0.30 & 0.83  & 0.22 & 0.27 & 0.81 & $P_{mbb}$ \\
4$_{11}$ & 6715 &  0.07 &-1.35  & 0.54 & 0.50 & 0.68 & $P_{mbb}-2,u$ \\
4$_{10}$ & 6773 & -0.28 & 3.35  & 0.32 & 0.32 & 0.72 & $P_{mbb},s$ \\
4$_{9}$  & 6809 & -0.59 &-0.47  & 0.38 & 0.70 & 0.47 & $dl$ \\
4$_{8}$  & 6875 &-0.37  & 0.13  & 0.13 & 0.29 & 0.34 & $mixed$ \\
4$_{7}$  & 6965 & 0.15  & 0.09  & 0.47 & 0.30 & 0.35 & $*,P_{mn}-2$\\
4$_{6}$  & 7017 &-0.08  & 1.18  & 0.30 & 0.27 & 0.27 & $*,P_{mbb}-2,s$ \\
4$_{5}$  & 7101 &-0.01  & 0.32  & 0.36 & 0.57 & 0.40 & $mixed$ \\
4$_{4}$  & 7168 &-0.81  & 0.19  & 0.27 & 0.78 & 0.26 & $P_{mn},u$ \\
4$_{3}$  & 7241 & 1.03  &-0.06  & 0.33 & 0.61 & 0.41 & $P_{mn}-1,s$ \\
4$_{2}$  & 7287 & 0.64  & 0.41  & 0.23 & 0.65 & 0.29 & $P_{mn}$ \\
4$_{1}$  & 7470 & 1.81  & 0.07  & 0.49 & 0.99 & 0.47 & $P_{mn},s$ \\
\end{tabular}
\end{ruledtabular}
\end{table}

The states for $P=4$ can be assigned in a similar fashion and the assignments
are shown in table \ref{table 2}. In particular, all the states
except 4$_{8}$ and 4$_{5}$ can be assigned in analogy with the states
in $P=3$. Thus the states 4$_{8}$ and 4$_{5}$ are classified as mixed.
In Fig.\ref{fig 2} states with large positive (4$_{10}$) and
large negative (4$_{14}$) $mbb$ velocities are shown. Note the
localization of the Husimis in the stable and unstable regions of
the phase space. 
The state space,
Husimi and partial velocity data for the mixed state 4$_{8}$ are 
shown in Fig.\ref{fig 2} as an
example. Extensive delocalization is seen for this state in both
the state space and the phase space.
The state 4$_{8}$ also happens to have the lowest IPR among the
states in $P=4$.
We note that for $P=4$ the maximum expected velocities for the $mn$
and the $mbb$ are $2.0$ and $4.4$ respectively.

A general observation regarding the assignments is that the lower
energy states of the polyads are dominated by $mbb$ whereas the higher end
states are influenced more by the $mn$ resonance. 
States in the middle of the energy regions exhibit
transitional character. Similar observations
were made in the dynamical assignments of the states in $P=8$ by
Jung {\it et al}\cite{jung}. Note that the assignments
made in this work are essentially based on the level velocities and the
IPRs. Thus our assignment does not invoke any explicit quantum
numbers based on periodic orbits. However it is clear that attaching
an approximate polyad and noting the signs on the level velocities is
providing dynamical information. For instance state 4$_{1}$ having
a large $\dot{x}_{mn} > 0$ implies that the state is influenced
by a stable periodic orbit of the $mn$ type. 
In particular this state would be
considered as a $C$ class state with the assignment $(l=4,t=0)$
in the notation of the previous work\cite{jung}. 
The approximate polyad $P_{mn}$ can be identified with the
longitudinal quantum number $l$.
A crucial point regarding the assignments using level velocities is worth
reiterating at this juncture. The advantage of using the level velocities
is based on strong correlation of the velocities
with the phase space nature of
the eigenstates. Such dynamical informations cannot be obtained by using
the IPRs alone or the various representations of the eigenstates.
For example, in $P=3$ the IPRs $L_{mn}(3_{3}) \approx L_{mn}(3_{4})$ and
one would correctly infer that both states are of $mn$ type. However
with $\dot{x}_{mn}(3_{4}) \approx -\dot{x}_{mn}(3_{3})$ it is possible
to provide a finer distinction between the two states ; 3$_{3}$ is
associated with a stable region of the phase space whereas 3$_{4}$ is
located in the unstable region of the phase space.

\section{Local frequency analysis: phase space diffusion and transport}

As mentioned before the Chirikov
analysis of DCO resonances based on the zeroth order classical
$H_{0}({\bf I})$ is not very accurate
due to strong resonant couplings of the various modes.
The surface of sections indicate that for polyads $P=3,4$ there is
considerable stochasticity even at the lowest energy. This implies that
the various resonances overlap strongly for these polyads and most
trajectories are chaotic. Despite the significant amount of chaos in
the phase space the Husimis of many eigenstates were seen to be rather
localized. Such observations have been made earlier in other studies and
it is well known that a largely chaotic phase space can still contain
structures that act as partial barriers to phase space transport.
The partial barriers can arise due to cantori as well as broken
seperatrices and existence of such structures can lead to
eigenstate localization. Moreover even in the case of strongly chaotic
systems eigenstates can get scarred by periodic orbits leading
to localization. In recent years it has been understood that such
partial barriers and periodic orbit scarring lead to 
statistically significant
deviations from the predictions of RMT.
Infact the level velocities, studied in the previous section, already encode
such localization information. 
From the IVR perspective, several studies\cite{mg20,mgtca}
have now established that the energy flow process is hierarchical and
that a hierarchical local random matrix (HLRM) approach is more appropriate
to understand IVR. 
The important factor is the local density of coupled
states and not the total density of states at a given energy which leads,
amongst other observations, to a power law decay of the survival
probability on intermediate timescales\cite{pow}. 

\begin{figure}
\includegraphics*[width=2in,height=3in]{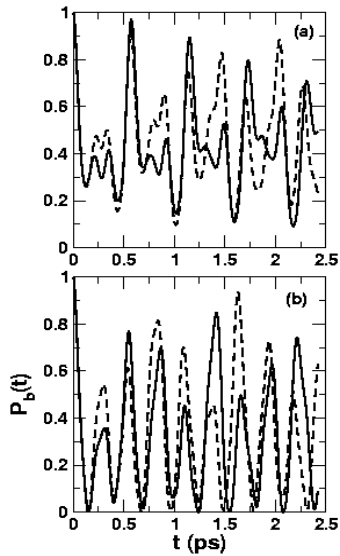}
\caption{Survival probabilities for the ZOBS (a) $|030\rangle$ and
(b) $|040\rangle$. Solid line is the exact Hamiltonian and dashed
lines are with $k_{nbb} = 0$.}
\label{fig 3}
\end{figure}

Based on the detailed work on 
classical phase space transport\cite{jdmrev} it is
tempting to conjecture that the hierarchical nature of IVR is intimately
connected to the various partial barriers in the phase space.
The important work\cite{lfa1,lfa2}
by Davis and Martens {\it et al} on OCS seems to support 
the conjecture. 
However, except for earlier works on OCS, there have not
been many efforts to identify partial barriers for transport in the
context of IVR in molecules. The importance of such an undertaking lies
in the fact that it might prove crucial in controlling IVR as well.
In DCO it has been established that 
the unimolecular decay is IVR limited\cite{renth,dcoexpt}.
Moreover, previous
studies\cite{renth,jung,keller} on DCO suggest that the Fermi 
resonance between the CO stretch and
DCO bend is unimportant. 
The approximate Chirikov analysis also indicates the muted role of
the $nbb$ Fermi resonance.
In Fig.\ref{fig 3} the survival probabilities for the zeroth order
bright states $|030\rangle$ and $|040\rangle$ in the absence of
the $nbb$ Fermi resonance are shown and compared to the full system.
It is clear from the figure that although the initial decay from the
bright states is controlled by the $1:1$ (for about 100 fs) the $nbb$ Fermi
resonance starts to play a role around 200 fs. 
Is it the case that the $nbb$ resonance does play a 
role, albeit in a secondary fashion, in the short time
tier structure? 
Why is the initial decay from $|040\rangle$ apparently complete by about
150 fs as opposed to those of $|030\rangle$ and $|050\rangle$? In other
words what makes the $1:1$ so effective for the bright state $|040\rangle$?
The answers to these questions are not straightforward
and in this work we will try to provide some approximate answers.
We believe that the LFA technique is capable of
addressing some of these crucial issues. 
Ahead a brief introduction is given to the technique of LFA and we
refer the reader to the work by Martens {\it et al}\cite{lfa1} 
and Arevalo {\it et al}\cite{are1,are2}
for more details.

The classical limit of the spectroscopic Hamiltonians are nonlinear,
multiresonant Hamiltonians which are functions of 
the zeroth order actions ${\bf I}$ and
the angles ${\bm \phi}$. The classical limit Hamiltonians have a typical
form $H({\bf I},{\bm \phi}) = H_{0}({\bf I})+
\sum_{j} V_{j}({\bf I},{\bm \phi})$ where the various resonant perturbations
are denoted by $V_{j}$. In the absence of the perturbations the system
is described by $H_{0}$ which is classically integrable. In such cases
the various nonlinear frequencies $\Omega({\bf I}) = 
\partial H_{0}/\partial {\bf I}$ are time-independent and one can define
a frequency map ${\bf I} \rightarrow \Omega({\bf I})$ which, if invertible,
can be used to characterize the phase space. The frequency map can be used
to characterize the phase space, thanks to
the KAM theorem, for near-integrable systems as well.
For a system like DCO there are three frequencies 
${\bf \Omega} \equiv (\Omega_{m},\Omega_{n},\Omega_{b})$
and it is useful 
to visualize the dynamics in the
frequency ratio space. 
However for non-quasiperiodic trajectories, associated with the full
nonintegrable Hamiltonian, the various frequencies $\Omega({\bf I})$
vary with time since increasing perturbation strengths lead to the breakdown
of KAM tori and widespread chaos.
Despite the time dependence of the frequencies it has been observed in many
systems that the frequencies can remain approximately constant over
times corresponding to many vibrational periods. Thus even non-quasiperiodic
trajectories can spend an appreciable amount of time in one part of the
phase space before passing on to another region of the phase space. This in
turn implies existence of long time correlations and gives rise to the
concept of dynamically significant regions of phase space.
The concept of local frequencies {\it i.e.,} time-varying frequencies
allows one to identify such dynamically significant regions of phase space.

Two main approaches are available to extract the local frequencies from
a classical trajectory. The first one proposed by 
Martens {\it et al}\cite{lfa1},
and a similar one by Laskar\cite{lask}, uses Fourier transforms of short time
segments of long trajectories. 
We are interested in the more recent work\cite{are1,are2}
by Arevalo and Wiggins wherein the local frequencies are obtained
by a continuous wavelet transform of the trajectories.
In this approach a general time dependent function $f(t)$ is expressed
in terms of the basis functions constructed as translations and dilations
of a mother wavelet\cite{wavelet}
\begin{equation}
\psi_{a,b}(t)=a^{-1/2} \psi \left(\frac{t-b}{a}\right)
\end{equation}
The coefficients of this expansion are given by the wavelet transform of
$f(t)$, defined as
\begin{equation}
L_{\psi}f(a,b) = a^{-1/2} \int_{-\infty}^{\infty} f(t)
\psi^{*}\left(\frac{t-b}{a}\right) dt
\end{equation}
for $a > 0$ and $b$ real.
The wavelet transform gives the local frequency of $f(t)$ over a small
interval of time around $t=b$ and inverse of the scale factor $a$ is
proportional to the frequency. A distinct advantage of the wavelet approach
is that the time window automatically narrows for high frequency and
widens for low frequency. The value of the local frequency at a given
time $t=b$ is extracted\cite{are1}
by determining the scale $a$ which maximizes
the modulus of the wavelet transform.
Throughout this work, as in Arevalo and Wiggins,
we will use the Morlet-Grossman mother wavelet
\begin{equation}
\psi(t) = \frac{1}{\sigma \sqrt{2 \pi}} e^{2 \pi i \lambda t}
e^{-t^{2}/2 \sigma^{2}}
\end{equation}
with $\lambda=1$ and $\sigma=2$. The parameters $\lambda,\sigma$ can
be tuned to improve the resolution. Within the wavelet approach the
local frequencies can be obtained even for chaotic trajectories. For
a strongly chaotic trajectory the frequencies will vary considerably
and one can measure the ``diffusion'' of the $k^{th}$
frequency by evaluating
\begin{equation}
d_{k}(T) = \frac{1}{T} \int_{0}^{T} dt |\Omega_{k}(t)-
\bar{\Omega}|
\end{equation}
where $\bar{\Omega}$ is the mean frequency and $T$ is the total
time to which the trajectory is propagated.
This measure of diffusion characterizes the unstable regions of
phase space and thus low diffusion implies either quasiperiodic
motion or extensive trapping near resonance zones (see below).
For our application to DCO we choose initial conditions  
for the trajectories in the following fashion. A uniform grid
is chosen in the state (zeroth order action) space $(I_{m},I_{n})$
with $I_{b}$ being fixed by the polyad conservation.
A particular choice of the angles $(\phi_{m},\phi_{n},\phi_{b})$
then provides initial conditions corresponding to a specific slice of
the phase space. 
The numerically integrated trajectories over certain
time interval $[0,T]$ are expressed as
$z_{k}(t)=\sqrt{2 I_{k}(t)} e^{i \phi_{k}(t)}$ for $k=m,n,b$.
The three local frequencies $\Omega_{k}(t)$ are extracted
by computing the maxima of the modulus of the wavelet transforms
of $z_{k}(t)$ {\it i.e.,} max$_{a}|L_{\psi}z_{k}(a,b)|$.
Note that the maxima are obtained with a fractional precision of $10^{-8}$.
Details regarding the numerical procedure to extract the maxima
can be found in the work of Arevalo {\it et al}\cite{are1}.

\begin{figure}
\includegraphics*[width=3in,height=3in]{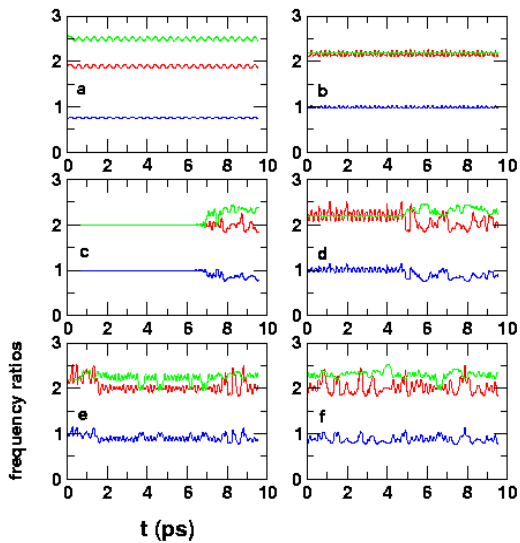}
\caption{The three frequency ratios for DCO for various trajectories
at energy $E = 5145$ cm$^{-1}$. Total time of propagation $T=10$ ps.
$\Omega_{m}/\Omega_{n}$ is shown in blue, $\Omega_{m}/\Omega_{b}$ is
shown in red and $\Omega_{n}/\Omega_{b}$ is in green. See text for
details on each trajectory.}
\label{fig 4}
\end{figure}

\begin{figure}
\includegraphics*[width=3in,height=3in]{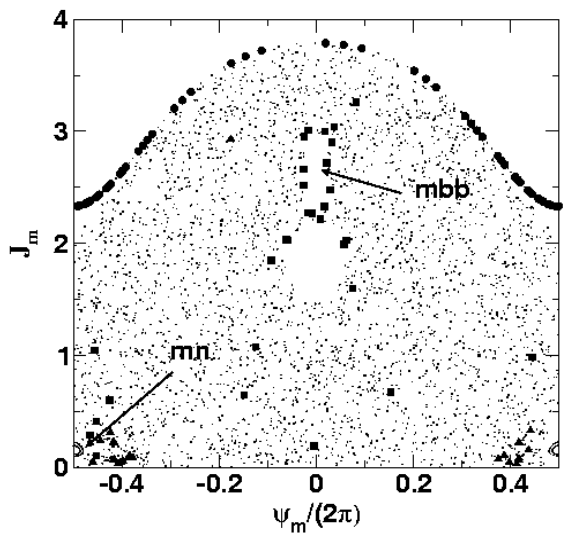}
\caption{Classical Poincar\'{e} surface of section at $E=5145$ cm$^{-1}$.
The variables are $J_{m} = I_{m}$ and $\psi_{m} = \phi_{m}-2 \phi_{b}$.
Selected trajectories (cf. Fig.\ref{fig 4})
are highlighted with filled circles (a),
triangles (d), and squares (e). The highlighted trajectories
are plotted for about 5 ps.}
\label{fig 5}
\end{figure}

A large number of trajectories for both $P=3$ and $P=4$ at various
energies were analysed and many distinct dynamical behaviours were
repeatedly seen.
In Fig.\ref{fig 4} we show, as examples of such dynamical patterns, 
six trajectories whose frequency ratios
were computed to about $T=10$ ps. Note that 10 ps is a fairly long time and
corresponds to about 500 CO vibrational periods. All of the six
trajectories are at an energy $E=5145$ cm$^{-1}$ which is energetically
close to the middle of the polyad $P=3$. 
The phase space structure is shown in Fig.\ref{fig 5} for reference and
one can clearly see large regions of stochasticity.
Figure (a) shows a typical quasiperiodic, nonresonant trajectory
which in this case represents a distorted local mode.
The corresponding trajectory is shown in the 
surface of section (circles in Fig.\ref{fig 5}) as well.
In Fig.\ref{fig 4}b a $1:1$ resonant regular trajectory corresponding to
the small island on the surface of section is shown. Notice the
locking of the relevant frequency ratio {\it i.e.,} $\Omega_{m}/\Omega_{n}
\approx 1$ for the entire duration. In Fig.\ref{fig 4}c a trajectory with
initial conditions very close to a periodic orbit of the full system is
shown. All the frequency ratios are locked perfectly for about 7 ps and later
show large diffusion. Interestingly in this case the $nbb$ locking
$\Omega_{n}/\Omega_{b}=2$ is clearly seen and thus any other initial condition
close to this would certainly involve the $nbb$ resonance to some extent.
In Fig.\ref{fig 4}d we show an example of a 
trajectory exhibiting resonance
trapping. This particular trajectory is trapped in the $1:1$ resonance
for nearly 5 ps and then drifts away. The corresponding trajectory is
plotted in the surface of section (triangles in Fig.\ref{fig 5})
and it is seen that this trapping
is due to the stickiness of the small $1:1$ island. The trajectory is chaotic
for longer times but the observed stickiness suggests that even tiny
regions of regularity in the phase space can sufficiently influence the
dynamics. Another example of a trapped trajectory is shown in
Fig.\ref{fig 4}e with the corresponding behaviour on the surface of section
shown in Fig.\ref{fig 5} (squares).
In this case the trajectory starts out with frequency locked in the
$1:1$ but drifts away quickly and gets trapped into the $\Omega_{m}/\Omega_{b}
\approx 2$ ratio for almost 6 ps. 
Finally in Fig.\ref{fig 4}f we show a typical
chaotic trajectory with large diffusion and no apparent long time
trapping in any of the fundamental resonances.
Two points are worth reiterating in the context of the examples shown
in Fig.\ref{fig 4} and Fig.\ref{fig 5}. 
The first is that all these various, and other, dynamical
behaviors are found at the same energy indicating the rich
structure inherent to the classical phase space. 
This energy is by no means special and indeed similar observations
hold for other energies in the polyad $P=3$ as well as different polyads.
Secondly, the overall phase
space might look chaotic on long time scales but the regular regions,
even those with phase space area $< \hbar$, lead to significant
stickiness\cite{stick,are2} and hence long time trappings in one or the other
significant resonance zone. This observation, we emphasize again, 
is not new and is
well known in the nonlinear dynamics community.
However in this work we are demonstrating this effect for a realistic
spectroscopic Hamiltonian with important consequences for
IVR in the molecule.

\begin{figure}
\includegraphics*[width=3in,height=3in]{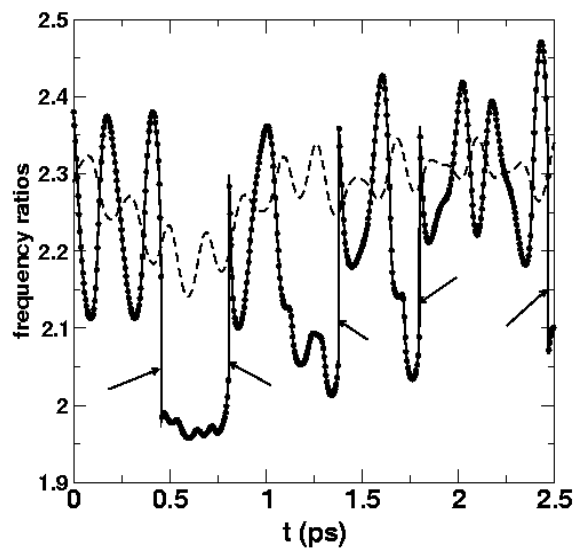}
\caption{A trajectory for $P=4$ exhibiting rapid frequency jumps.
For clarity the ratios $\Omega_{m}/\Omega_{b}$ (solid line with
points) and $\Omega_{n}/\Omega_{b}$ (dashed) are shown. Note that
the jumps (indicated by arrows)
take the trajectory from being near $mbb$ resonance to
being near the $mn$ resonance. Only a 2.5 ps portion of the
15 ps trajectory is shown. The jumps occur over the entire time span.}
\label{fig 6}
\end{figure}

\begin{figure*}[!]
\includegraphics*[width=5.0in,height=5.0in]{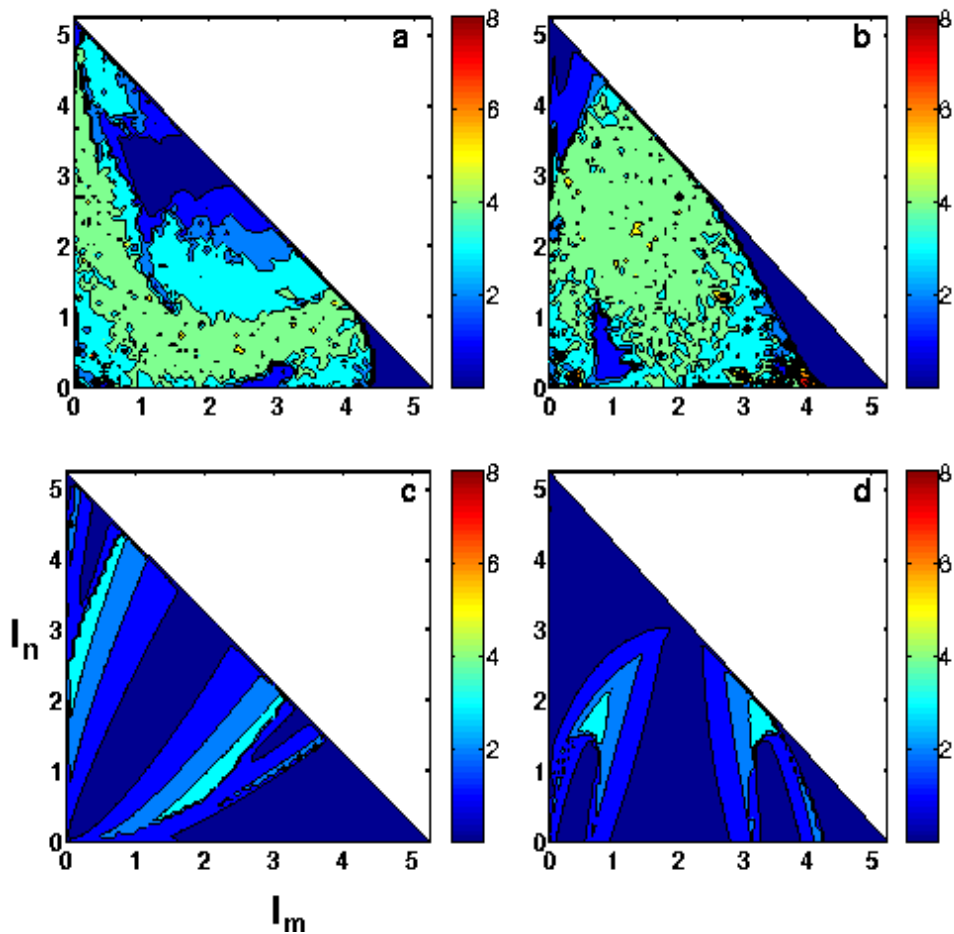}
\caption{Total diffusion (ps$^{-1}$) in the state space for $P=4$ computed
for $T=10$ ps. In (a) and (b)
the phase space slices are $(0,0,0)$ and $(0,\pi,0)$ respectively.
The integrable subsystems are shown for reference in (c) $mn$ and
(d) $mbb$ respectively. In (c) and (d) the resonance zones can be
seen clearly. In (a) and (d) there are no isolated resonances and
the system is strongly coupled.}
\label{fig 7}
\end{figure*}

A further dynamical effect that occurs in DCO has to do with rapid
jumps in the frequency ratio space. An example for $P=4$ is shown
in the Fig.\ref{fig 6} where the ratio $\Omega_{m}/\Omega_{b}$ is
undergoing rapid jumps between otherwise smooth variations. Note that
the ratio $\Omega_{n}/\Omega_{b}$ is fairly smooth and does not
exhibit the rapid jumps. The jumps in the frequency ratio happen
from the trajectory being near the $mbb$ resonance to the trajectory
being near the $mn$ resonance. Such rapid jumps have been observed
by Arevalo in OCS as well as the planar 3-body problem\cite{are2}.
In this instance the jumps seem to correspond to heteroclinic
orbits that connect the $mbb$ and the $mn$ resonance zones. However
confirming this involves computing invariant manifolds and their
intersections which is far from easy. Nevertheless, such rapid
resonance transitions could have important implications for the
nature of IVR in DCO.

To indicate the strongly coupled nature of the DCO dynamics
we show the total diffusion for $P=4$ over the
state space in Fig.\ref{fig 7}. In Fig.\ref{fig 7}a the diffusion
is shown for the phase space slice $\phi_{m}=\phi_{n}=\phi_{b}=0$.
For reference Figs.\ref{fig 7}c,d show the 
diffusion data for integrable $mn$ and
$mbb$ subsystems. It is easy to see the $mn$ and $mbb$ resonance zones
in the integrable single resonance cases. Comparing to Fig.\ref{fig 7}a
it is clear that the resonances have overlapped strongly. 
Further in Fig.\ref{fig 7}b we show the diffusion for the phase
space slice $\phi_{m}=0=\phi_{b},\phi_{n}=\pi$. Note that the
diffusions for the two slices of phase space are dramatically
different. This implies that it is not possible to understand the
intramolecular dynamics of DCO based on $H_{0}$ alone.
A key observation here is that the diffusions in state space show
significant heterogeneity. For instance in Fig.\ref{fig 7}a,b we have
regions of similar diffusion strengths seperated by a region of
larger diffusion strengths. Thus if one imagines going along
a specific $mn$ or $mbb$ polyad in state space then alternating
diffusion strengths are encountered. Similarly along a specific energy
contour the diffusion values can vary considerably. This leads to
trappings of the classical trajectory and hence bottlenecks to 
energy flow. For instance in Fig.\ref{fig 7}b the
region of low diffusion around $(I_{m},I_{n})\approx (1.0,0.5)$
is due to the existence of a periodic orbit of
the full system. In the semiclassical limit the heterogeneous nature
of the state space diffusion can have interesting consequences for
eigenstate localizations.

Temps {\it et al.} discuss\cite{renth} the differences between choosing
a CO-stretch bright state (experimental) versus the DCO bend
bright states. It was shown that the DCO bend states undergo much
faster unimolecular decay. In order to gain dynamical insights
into their findings the various bright states were analyzed from the 
LFA perspective. Several initial conditions corresponding to the
bright states {\it i.e.,} $E = E_{ZOBS}$ were generated with
the actions
$(I_{m},I_{n},I_{b})$ corresponding to the bright state of interest
and randomly choosing the angle $(\phi_{m},\phi_{n},\phi_{b})$ variables.
The various dynamical quantities of interest are then averaged
over the angles. This is done for two main reasons. First, although
the correspondence between actions and quantum numbers is well known
a similar correspondence for angles is not very clear. In the
deep semiclassical limit one expects a correspondence but the problem
at hand is far from such a limit. Second observation is that the
quantum dynamics associated with a particular ZOBS, semiclassically
speaking, would naturally involve a family of classical trajectories which
are modeled here with the random angle distributions.
The fact that the diffusion of the frequency ratios are strongly
dependent on the choice of the phase space slice further support
the procedure of angle averaging.
The quantities of interest are 
number of crossings per unit time of the two 
main resonances (denoted $\langle c_{mn},c_{mbb} \rangle/T$),
total time spent near the 
resonances (denoted $\langle \tau_{mn},\tau_{mbb} \rangle$), 
longest locking time in a given resonance (denoted $\langle \tau_{m}
\rangle$),
and the diffusions $d(T)$. In this study we have averaged over
1000 trajectories with the angles chosen such that 
$E = E_{ZOBS} \pm 1.0$ cm$^{-1}$ and each trajectory was propagated
to $T = 10$ ps.

\begin{figure}
\includegraphics*[width=3in,height=3in]{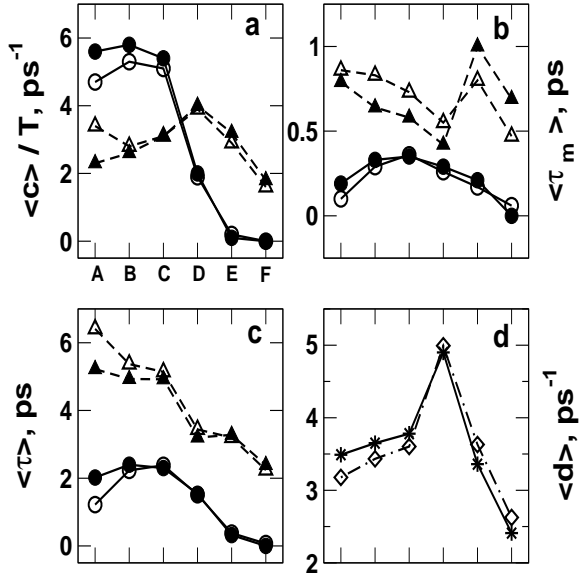}
\caption{Average crossings per unit time (a), average longest
locking time (b), total time near a specific resonance (c) and
diffusion of the frequencies (d) are shown in this figure for
six ZOBS. A, B, C, D, E, F correspond to the bright states
$|030\rangle, |040\rangle, |050\rangle, |006\rangle, |008\rangle$, and
$|0010\rangle$ respectively. Averaging is done over 1000 trajectories
and $T = 10$ ps. In (a), (b), and (c) the filled circles
and triangles correspond to the $mn$ and $mbb$ resonances
respectively while the open symbols are for the case
with $k_{nbb}=0$. See text for details.}
\label{fig 8}
\end{figure}

In Fig.\ref{fig 8} we show the various dynamical quantities of interest for
three of the experimental CO stretch bright states and three of
the bend only bright states corresponding to the polyads $P=3,4,5$.
In Fig.\ref{fig 8}a the number of resonance crossings per unit time are
shown. It is immediately clear that the $mn$ resonance plays a dominant
role for the CO stretch bright states and that the $mbb$ resonance
is key for the bend bright states. Notice that for the $|008\rangle$
and the $|0010\rangle$ bright states the 'flux' across $mn$ is
almost zero. On the other hand despite the dominance of $mn$ flux
for the CO stetch ZOBS, the $mbb$ flux is not negligible. This already
provides a hint to the drastically different IVR dynamics from the two
classes of ZOBS with $|003\rangle$ being an exception.
Fig.\ref{fig 8}b shows the longest locking time, on average, for the different
bright states. Interestingly even though the $mn$ flux is larger
than the $mbb$ flux for the CO stretch states the $\tau_{m}$'s show
the opposite trend. The total time spent near a particular 
resonance as shown in Fig.\ref{fig 8}c also indicates that the IVR from
CO stretch states is strongly effected by the $mbb$ resonance. The diffusions
in Fig.\ref{fig 8}d show marginal differences. A clear picture emerges for
the IVR dynamics associated with the CO stretch bright states. 
Initially the $mn$ resonance transfers energy to the DC stretch (dissociation
mode) but prior to a substantial buildup of excitation in the DC stretch
the energy is ``siphoned off'' into the bend mode via the $mbb$ resonance.
This process happens for a fairly long time leading to a highly
nonstatistical nature of the dynamics and unimolecular decay of DCO.
On the other hand for the bend bright states corresponding to $P=4,5$
the initial IVR dynamics is dominated by the $mbb$ resonance. However the
$mn$ resonance is unable to compete and this results in unhindered transfer
of energy to the dissociative mode via the $mbb$ coupling. Exception
to the nature of bend ZOBS is clearly seen in the case of $|006\rangle$
wherein the $mn$ resonance does compete and leads to transfer of
energy from the dissociative mode to the CO stretch mode.
An important observation is that the diffusion of frequencies computed
here is the lowest for the ZOBS $|0010\rangle$ which undergoes the
fastest unimolecular decay. This is not a contradiction since large
difusion corresponds to many hops and consequent locking leading to
incomplete flow of energy into the dissociative mode. For example the
largest diffusion is predicted for the ZOBS $|006\rangle$ which is
consistent with the fact that the times spent near the two main resonances
are similar as compared to the other states. Thus $|006\rangle$ should
decay the slowest which is confirmed by computing the associated
survival probability (cf. figure 3(b) in Ref.~\onlinecite{renth}).
It is interesting to observe that the diffusions for the CO stretch
bright states are not very different whereas in the case of the
bend bright states there is a large variation. This demonstrates the
mode-specificity of the intramolecular dynamics of DCO.

In Fig.\ref{fig 8} the effect of setting $k_{nbb}=0$ is shown. It is
clear that the largest effect of the $nbb$ Fermi resonance is for
the CO stretch bright states $|030\rangle$ and $|040\rangle$.
In the absence of $nbb$ the $mn$ flux is decreased and the
$mbb$ flux is increased with larger time being spent, on the average,
near the $mbb$ Fermi resonance. This indicates that the CO stretch
bright states would lead to smaller rates of unimolecular decay in the
absence of the $nbb$ resonance. This is true for the bend bright states
as well although the effect is very small. Thus the
role of $nbb$ Fermi resonance is not negligible for the lower polyads.
As a final observation we point out that amongst the CO stretch
bright states there is a sharp
increase, on going from $|030\rangle$ to the $|040\rangle$, in
the average residence time and the average longest locking time
for the $mn$ resonance. This perhaps explains the efficiency of
the $mn$ resonance in $P=4$ but at this stage it is conjectural.

\section{Conclusions}

A long term goal of our research is to come up
with techniques that would allow fairly detailed classical-quantum
correspondence studies of highly excited states in systems with
three degrees of freedom or more. 
To this end two dimensionality independent techiniques,
parametric variations and local frequency analysis, have been
employed in this study. We have demonstrated the utility of both
the techniques for understanding the
intramolecular dynamics of DCO. 
The successes of both the techniques, in dynamical
assignments of states and establishing the
nonstatistical nature of the dynamics, given the strongly coupled
nature of the system (coupling strengths comparable to the mean
level spacing) is quite encouraging. 
We conclude with brief
comments on the techniques and future outlook.

The level velocity approach, used to dynamically assign the eigenstates, is
certainly capable of identifying strongly localized states. The
close connections between the level velocity of an eigenstate
and its corresponding phase space nature makes this approach 
attractive. Previous application\cite{par1} to a mixed 2-mode system
demonstrated the utility of the level velocities for dynamical
assignments of the highly excited states. In this work we have
shown that the method is useful even in the presence of significant
chaos. Although avoided crossings lead to difficulties, and this is
not surprising, it is debatable wether the resulting mixed states
can be or should be assigned. Demixing the states invovled in 
avoided crossings and then analyzing them 
can be done within this approach as in any other
approach. However the procedure of demixing, especially in
multistate or broad avoided crossings, should not be considered as
an assignment. This work shows that analyzing the level velocities
in conjunction with measures like IPRs can be a fairly useful way
of dynamically assigning highly excited states. At this stage the 
effects of various bifurcations on the level velocity spectrum is
not known. Interestingly in the case of DCO, despite substantial
change in the phase space structure with changing energies, the
level velocity corresponding to a state localized about stable regions
is very close to the classical estimate. 
This has been observed before\cite{par1} in
the case of HCP as well. This suggests a very detailed classical-quantum
correspondence and needs to be studied further. In order to make this
approach more useful it is important to obtain some bounds on the
level velocities in the mixed and strongly chaotic situations. Thus a
semiclassical theory for the velocities in the chaotic regimes needs
to be developed and work is in progress in our group along these lines.
Note that the distribution of the velocities has been found to be
gaussian in the strong chaos limit\cite{haake}. 
The general technique of parametric
variations has a wide range of applications and level curvatures have
also been studied\cite{haake}. 
It remains to be seen if the level curvatures
have any utility in understanding highly excited states and IVR.
For example conductance of disordered systems have been related to
level curvatures\cite{haake,ipr,mont} and given 
the tantalizing similarities between the
example mentioned and IVR it seems worthwhile to 
investigate the utility of level curvatures in molecular systems.

The LFA technique is able to provide detailed phase space information
in a strongly coupled system like DCO. In particular the time-frequency
analysis revealed the existence of trajectories trapped near
important resonance zones. In an effective spectroscopic Hamiltonian
the key resonances are easy to expect since the
fit already builds in the main resonant perturbations
but LFA is able to provide
information on the competing roles of these various resonances. An
example is the clear differences between the CO-stretch bright state
and DCO bend bright state dynamics for DCO. LFA is also able to 
highlight the role of weak resonances like the $nbb$ Fermi resonance.
As shown in this work and in the previous works the wavelet based
approach to LFA is useful in analyzing effective Hamiltonians and
various dynamical quantities can be determined to understand IVR.
LFA clearly shows that despite significant
chaos the DCO phase space is far from
the RMT regime. 
In this paper it is shown that even for a strongly coupled system
like DCO the state space diffusion has a lot of structure with
regions of high and low diffusions interspersed thorughout.
The diffusion of frequencies provides clear insights
into the dynamics at a particular polyad and/or a specific energy.
This information can then be used to determine the nature of
the 'random walk' in action space for strongly coupled systems
and thus provide more insights into certain recent conjectures\cite{pow}
regarding the intermediate time behavior of survival probabilities.
Although not shown here most of the trajectories exhibit across-resonance
diffusion which is quite fast and implicates higher order secondary 
resonances as well\cite{lask,hall}. 
The rapid frequency jumps which perhaps correspond to
heteroclinic orbits joining unstable periodic orbits have not
been investigated in detail. Surely the rapid jumps from one
resonance to another have important consequences for IVR 
and needs further study.
We also note that there is not much of an
Arnol'd web for DCO for $P=3,4$ in contrast\cite{are1} to H$_{2}$O.
DCO being effectively two dimensional implies that there is no
transport along resonance lines and thus many of the interesting
conjectures put forward\cite{lfa1,lfa2} by 
Martens, Davis, and Ezra remain to be
tested. Currently work is in progress for true three degree of
freedom systems.

As a final note we mention that both techniques used in the current study
probe localiztion characteristics. The parametric response is probing
eigenstate localization and the LFA is probing localization of phase
space transport due to partial barriers and broken seperatrices. 
Localization explicitly implies nonstatistical behavior 
and thus deviations from RMT and incomplete IVR in the system.
Evidence exists\cite{rad,wyatprl} that 
the quantized versions of the partial
barriers can be more restrictive than the classical analogs.
A interesting link between these two approaches is provided by
the recently proposed intensity-velocity correlator\cite{corr1,corr2,corr3}. 
Detailed studies of true three degrees of freedom system via
the intensity-velocity correlator can lead to a better understanding
of the IVR dynamics.

\section{Acknowledgements}
It is a pleasure to acknowledge useful discussions with Dr. L. V. Arevalo
regarding the wavelet based approach to time-frequency analysis.
We are also grateful to Prof. Greg Ezra for sending us a copy of
the unpublished work Ref.\onlinecite{lfa2}.
This work is supported by funds from the Department of Science and
Technology, India.

\end{document}